\documentclass[journal]{IEEEtran}
\usepackage{graphicx}
\usepackage{multirow}
\usepackage{hyperref}
\usepackage{url}
\usepackage{placeins}

\title{Making a Case for Federated Learning in the Internet of Vehicles and Intelligent Transportation Systems}

\author{Dimitrios Michael~Manias and
        Abdallah~Shami \thanks{Dimitrios Michael Manias and Abdallah Shami are with the Department of Electrical and Computer Engineering at the University of Western Ontario e-mail: \{dmanias3, Abdallah.Shami\}  @uwo.ca}}

\begin{document}

\maketitle

\begin{abstract}
With the incoming introduction of 5G networks and the advancement in technologies, such as Network Function Virtualization and Software Defined Networking, new and emerging networking technologies and use cases are taking shape. One such technology is the Internet of Vehicles (IoV), which describes an interconnected system of vehicles and infrastructure. Coupled with recent developments in artificial intelligence and machine learning, the IoV is transformed into an Intelligent Transportation System (ITS). There are, however, several operational considerations that hinder the adoption of ITS systems, including scalability, high availability, and data privacy. To address these challenges, Federated Learning, a collaborative and distributed intelligence technique, is suggested. Through an ITS case study, the ability of a federated model deployed on roadside infrastructure throughout the network to recover from faults by leveraging group intelligence while reducing recovery time and restoring acceptable system performance is highlighted. With a multitude of use cases and benefits, Federated Learning is a key enabler for ITS and is poised to achieve widespread implementation in 5G and beyond networks and applications. 
\end{abstract}

\maketitle

\section{Introduction}
With recent advancements in networking technologies such as Mobile Edge Computing (MEC) and Network Function Virtualization (NFV) and the incoming introduction of Fifth Generation (5G) networks, a multitude of new applications and use cases are being realized, such as the Internet of Vehicles (IoV). These up-and-coming use cases require stringent Quality of Service (QoS) guarantees and strict Service Level Agreements (SLAs) to ensure network properties such as scalability, flexibility, elasticity, high availability, and performance; however, these properties cannot be realized to their full potential without the use of intelligence techniques such as Machine Learning (ML) and Advanced Analytics (AA). By leveraging the enabling networking technologies to create the IoV and combining it with intelligence techniques, the IoV transforms into an Intelligent Transportation System (ITS).

\subsection{Mobile Edge Computing}
MEC is a technology that pushes cloud computing services to the network edge. By utilizing the network edge, several benefits arise, including ultra-low latency, high bandwidth, and real-time applications. Due to these benefits, MEC has been highlighted as an enabling technology for next-generation networking and applications. Some examples of use cases stemming from MEC implementations include video analytics, location services, IoV and augmented reality \cite{AA1}. Another major advantage of MEC implementations is the reduced traffic experienced on core networks, something which is essential given the growing network connectivity demand experienced globally by Network Service Providers (NSPs). When considering use cases such as the IoV, real-time, ultra-low latency processing available at the edge of the network is essential for critical services, including collision avoidance and infotainment services, such as virtual and augmented reality. However, MEC itself requires the virtualization of network infrastructure to enable cloud resource utilization at the network edge. 

\subsection{Network Function Virtualization}
NFV technology was proposed by the European Technical Standards Institute in 2012 and entails the abstraction of network functions from dedicated hardware \cite{BB1}. Once abstracted, the network functions are converted to Virtual Network Functions (VNFs), which are software-based applications that run on universal hardware such as data center servers and edge servers. Through NFV technology, several benefits can be realized, such as reduced capital and operational expenditures, improved network performance and operation, and improved network health \cite{CC1}. When considering NFV technology, one of the greatest challenges is the management and orchestration (MANO) of VNFs, which includes tasks such as the placement of VNFs on network servers, VNF scaling and VNF migration. The increasing complexity of networks, coupled with the NP-hard computational complexity of these problems \cite{DD1}, has led NSPs to consider alternate approaches to address NFV MANO. 

\subsection{Intelligence and NFV}
Recently, the use of intelligence techniques such as ML and AA have been increasingly popular when considering NFV MANO functionalities. This increase in popularity is attributed in part to a major paradigm shift from analytical network modelling to data-driven network modelling. With the generation of increasing amounts of network data, NSPs are beginning to adopt intelligence technologies that leverage the previously untapped data and extract meaningful information. There are several benefits associated with the use of intelligence and the adoption of data-driven modelling in NFV MANO. Firstly, since network complexity is increasing, analytical system modelling becomes increasingly difficult; by modelling the system directly from the generated data, NSPs can get accurate system models without the need to describe the system mathematically in its entirety. Additionally, in the case of NFV MANO functionalities ({\it i.e.,} VNF Placement), intelligence can be used to learn from past optimal VNF placements and predict future placements in real-time. This ability to predict optimal placements enables real-time optimal decision making, something which was previously not possible due to the complexity of optimization problem formulations and the static nature of near-optimal heuristic solutions. Additionally, the use of intelligence enables a plethora of new and innovative functionalities such as traffic, demand, and latency prediction and forecasting, which can be used to optimize scaling operations. Additionally, one of the main functionalities of 5G and beyond networks is automation in the form of self-healing, self-configuration, and self-optimization; all of these functionalities require extensive use of intelligence for forecasting and prediction thereby making intelligence an essential requirement of all 5G and beyond networks. 

\subsection{Intelligent Transportation Systems}
The combination of MEC, NFV, and intelligence sets the stage for several emerging use cases, most notably, ITS  \cite{RV1}. The ITS framework envisions a system of connected vehicles communicating with each other and with Intelligent Infrastructure (II) using dedicated short-range communications. Through ITS connectivity, Vehicular Clients (VCs) will have access to multiple types of services, including traffic, emergency, and infotainment \cite{RV1}. Considering the use of MEC in this system, these services will be provided through NSPs by means of VNFs placed close to the user at the network edge. To enable this, several Roadside Units (RSUs), acting as edge servers, will be placed along roads. These units will have the capability of collecting information such as traffic and weather conditions, locally processing that data, and sending it to various entities within the ITS system. Figure 1 shows a basic ITS system and highlights the various entities previously mentioned (1 – RSU, 2 – VC, 3 – II). Figure 1 also highlights the presence of entities capable of performing local intelligence, marked with the gears.

\begin{figure}
\centerline{\includegraphics[width=16.5pc]{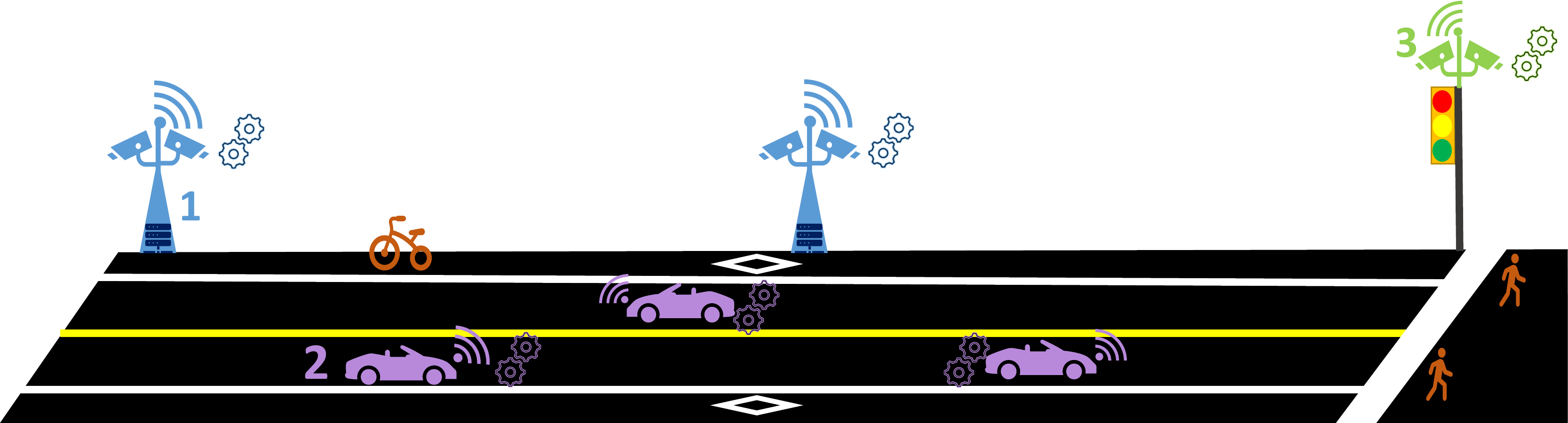}}
\caption{Basic Intelligent Transportation System Overview}
\end{figure}

\subsection{The Challenges of Machine Learning Implementation in ITS Systems}
There are several challenges regarding the implementation of ML in highly dynamic environments such as the IoV and ITS. These challenges are categorized into four major groupings, system complexity, model performance, privacy, and data management \cite{RV2}. System complexity is a major challenge when considering ITS as it is a very volatile environment; while roadside infrastructure may be constant in nature, vehicular clients are continually entering and leaving the system. This volatility presents a unique challenge regarding ML implementation as the operational domain is continuously changing, something which is not easily handled by traditional ML \cite{RV2}. The shifting operational domain leads to the second major challenge being model performance. As the functional domain changes, the performance of the model is severely impacted. Models using static local intelligence cannot adapt to changing environments and can become ineffective as their performance is severely degraded. Considering the criticality of an ITS system, the safeguard of human life is paramount. Any level of compromise in the system ranging from its infrastructure to its data, can endanger public safety. Finally, with an increasing number of network nodes with processing capabilities, the management of data becomes increasingly critical. Since the roadside infrastructure will have limited resource capacity, special consideration must be made regarding the efficient storage of data. Since the resource capacities of the roadside infrastructure will not allow for the storage of extensive data sets, the training phase of traditional localized ML techniques can be compromised due to the lack of sufficient data. 

\subsection{Why Federated Learning?}
When considering possible methods of advanced intelligence applied to an ITS, Reinforcement Learning (RL) and Federated Learning (FL) are two standout options. RL is an intelligence method capable of learning complex policy decisions in a dynamic environment and has the ability to adapt to a changing domain through continual and experiential learning \cite{RV1}. However, the success of the computationally intensive training phase of RL is greatly dependent on its simulated training environment, the design of its reward function, and the tuning of its hyperparameters \cite{RV1}. Considering an ITS with multiple light-weight points of presence having limited processing and storage capabilities, the implementation of RL becomes increasingly challenging. \par
When considering an ITS, security and privacy are paramount; this applies both to vehicles and the data they generate. Since not all intelligence is created equal, a technique that maximizes data privacy while still meeting the required performance is essential to the feasibility of such a system. Furthermore, when considering the system environment, a distributed and communication-efficient intelligence method is required. Additionally, a form of intelligence that can overcome faults and failures quickly, thereby ensuring system resilience and service continuity, is a major contender for implementation. To simultaneously address privacy concerns while ensuring a resilient and intelligent system capable of excellent performance, we make a case for FL in the IoV and ITS. The remainder of this paper is organized as follows. Section II Outlines FL, its advantages, and its benefits. Section III outlines the applications of FL and showcases possible use cases. Finally, Section IV summarizes and concludes the paper.

\section{Federated Learning}

FL is a key enabling technology for many emerging use cases. The following section will outline the main aspects of this technology, its benefits, and present a case study illustrating its effectiveness. Table 1 summarizes the various benefits provided by Federated Learning implementations, as outlined throughout this section.

\begin{table}[h]
\caption{Federated Learning Benefit Summary}
\begin{tabular}{|p{2cm}| p{6cm}|}
\hline 
Benefit & Description\\
\hline 
Collaborative Learning & Enables collaboration between multiple entities for intelligence.\\
\hline 
Distributed Learning & Enables intelligence across distributed nodes and enables intelligence
in new and emerging use cases.\\
\hline 
Decentralized Learning & Local nodes can train their own models and can operate independently
without the aggregation agent in failure scenarios.\\
\hline 
Data Availability & Allows the leveraging insights obtained from massive amounts of data
without the need for locally hosting the data.\\
\hline 
Data Privacy & Data is collected and processed at local nodes and not shared globally.\\
\hline 
Scalability & New nodes can easily be integrated into existing systems.\\
\hline 
Fault Recovery & Offers a rapid recovery scheme for nodes experiencing faults.\\
\hline 
High Availability & Offers system continuity during failure scenarios.\\
\hline 
Communication Efficiency & Model updates are sent instead of complete models or data.\\
\hline 
\end{tabular}

\end{table}

\subsection{What is Federated Learning}

FL is a machine learning technique first proposed by Google in 2017 as a way of providing decentralized and collaborative learning across distributed nodes \cite{EE1}. Initially, FL was considered for applications relating to smartphones such as text prediction; however, since then, it has been used in fields such as medicine and image recognition. The FL architecture consists of multiple federated nodes and an aggregator agent. Initially, a global model is created by the aggregator agent and is distributed to all the federated nodes. Once received, the nodes begin by training the model on their locally stored data. Since each node is responsible for the collection and/or storage of its own data, each federated node possesses a unique training data set. Once the federated nodes have trained their models for several iterations (pre-defined in the aggregation scheme), the nodes compare the initial global model received to their locally trained model. An update is generated by each federated node, whereby the results of the comparison are stored. It must be noted that this update does not include the locally trained model itself; instead, it lists the discrepancies between the global and local models, thereby highlighting the changes made during the training process. The aggregator agent then samples the nodes according to the aggregation scheme and collected the model updates. Once collected, the updates are used to create a new global model that is distributed to the nodes, and the process repeats itself.  Figure 2 outlines the FL process  at a high level. However, to fully appreciate the entirety of the FL process, a more granular analysis of the federated nodes and the aggregator agent is required.

\begin{figure*}[h]
\centerline{\includegraphics[width=25pc]{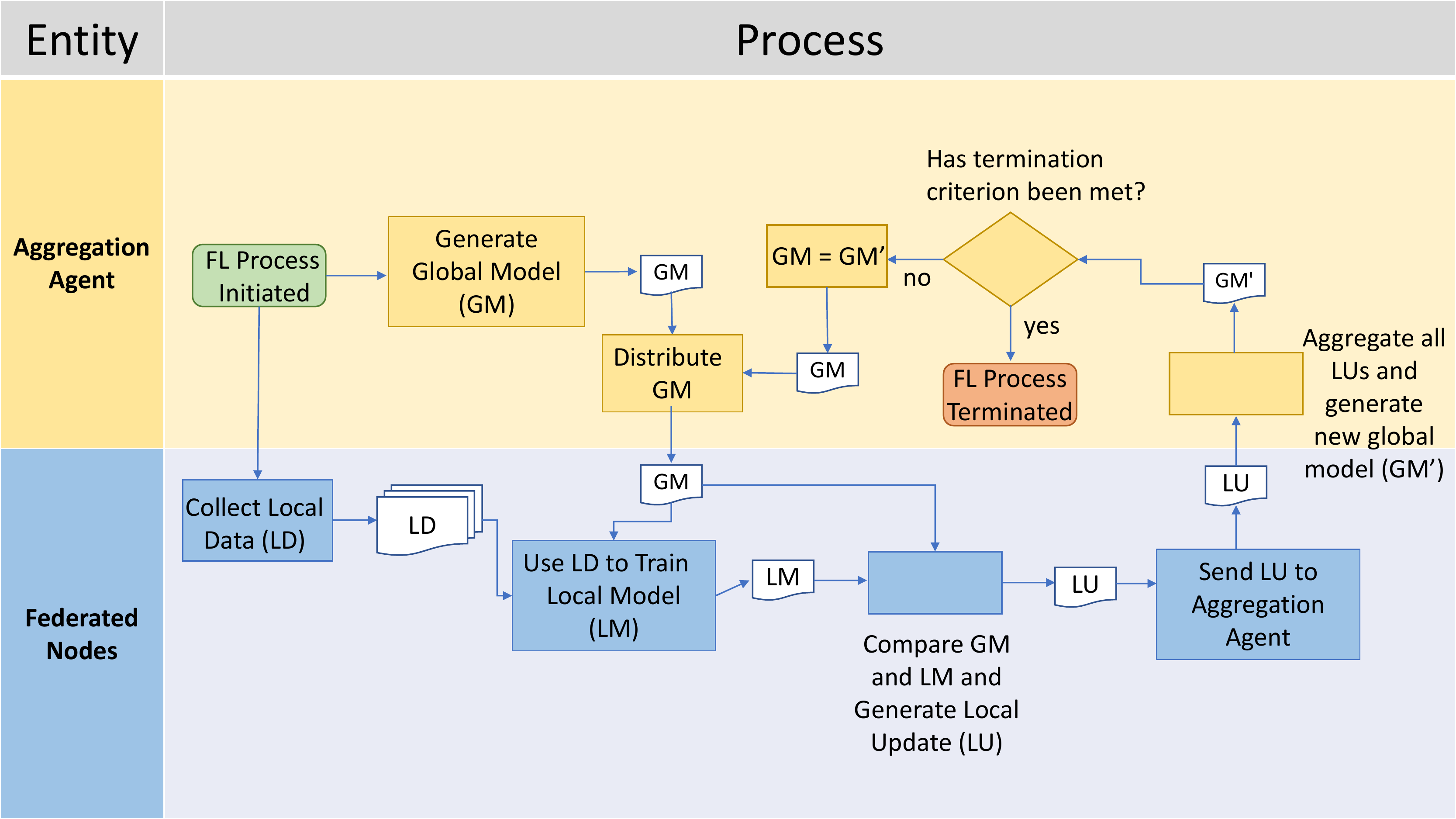}}
\caption{Federated Learning High-Level Process Map}
\end{figure*}

\subsection{Federated Nodes}

Federated Nodes are unique and versatile entities capable of data collection and processing, model training, and network communication. When considering federated nodes in IoV and ITS scenarios, the main entities which can be classified as such are VCs, RSUs, and II. In the envisioned ITS scenario, these entities will have various sensors and will be gathering a multitude of data. Each of these entities will have processing and communication capabilities.

\subsection{Aggregator Agent}

\begin{figure*}[h]
\centerline{\includegraphics[width=25pc]{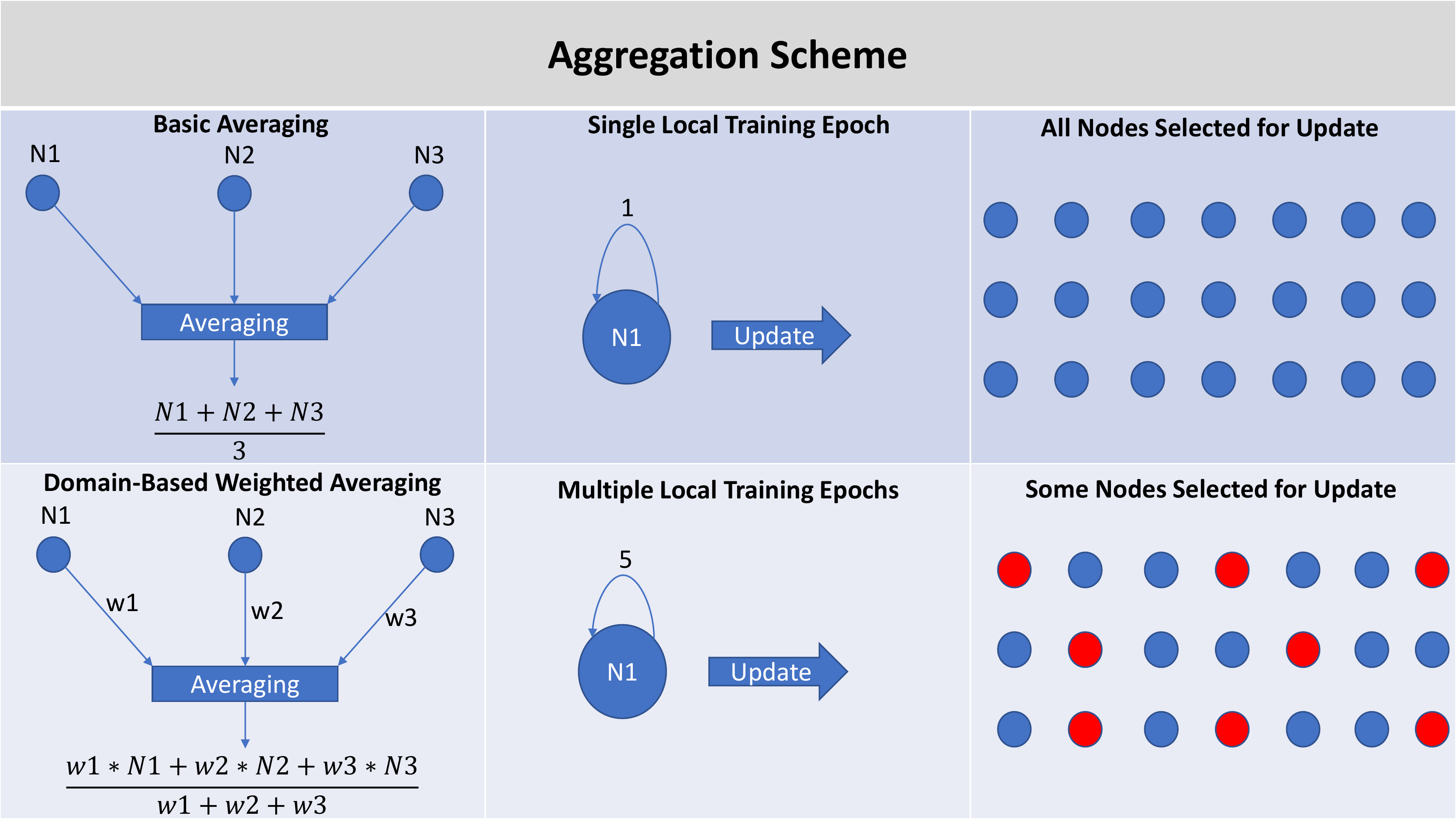}}
\caption{Aggregation Agent Scheme Considerations}
\end{figure*}

The aggregator agent is the main orchestrator behind the FL training process and is responsible for determining how often and how many nodes will be contributing to the global model update based on the aggregation scheme. Developing an aggregation scheme is the main challenge associated with FL as it directly affects the performance of FL process.   \par

When considering the aggregation scheme, the first decision to be taken is how the model updates received from each of the nodes will be consolidated into a single update used to generate the new global model. Traditionally, averaging has been used to combine all the model updates; however, increasingly, new strategies are emerging which take into consideration individual properties of the federated nodes, including resource availability and node criticality. By creating a more refined update aggregation strategy that captures domain-specific information, model performance can be greatly improved.   \par

The second consideration which must be made is the frequency of model updates. This frequency is a joint consideration between the number of local training iterations before the update and the time required to locally train the model. This consideration brings to light one of the main attributes of FL data; traditionally, most ML models assume that a dataset is independent and identically distributed (i.i.d.); however, this assumption is invalid for FL as it operates under the assumption that the data is non-i.i.d. When thinking about this intuitively, especially considering an IoV scenario that the data collected will not follow the i.i.d. assumption. Consider a scenario where a group of federated nodes are responsible for training object detection models at given intersections; the busier intersections will have more local data compared to those which are located in low-traffic areas, thereby creating a non-identically distributed scenario. Going back to the aggregation scheme, nodes with more local data will require a longer time to train their models and generate the update. This imbalance in data and, therefore, an imbalance in training time must be taken into consideration when selecting the frequency of model updates. \par

The final consideration, which must be made regarding the aggregation scheme, is the sampling strategy. Due to the nature of FL, especially in applications such as the IoV, all nodes will be gathering different data. Performing a model update using an insufficient amount of data is a futile task as it will not bring an overall benefit to the global model.  Conversely, waiting for a node with an excessive amount of data to finish its training can hold back the training process and drastically reduce the speed at which the global model is trained. Furthermore, due to the dynamic nature of the network supporting MEC implementations, there could be a high volume of traffic resulting in higher communication latency at a given node; using resources to communicate model updates at such a time can further burden the network and have adverse effects on QoS requirements and service delivery.  Finally, in the case of a fault or failure, a particular node can go offline; waiting for a model update at this time would be exclusively dependent on the extent of the fault and the ability to recover from the fault, and can greatly reduce the global model’s ability to progress in training. Taking these scenarios into consideration, a sampling of nodes must be conducted to determine which nodes will provide the greatest benefit to the current global model through their updates. This decision is domain-specific as in certain situations, the nodes with the greatest amount of data provide the most benefit, whereas, in other situations, the nodes with a moderate amount of data can speed up the training process and converge to a solution faster. Figure 3 outlines the various aspects of the aggregation scheme previously mentioned.

\subsection{What are the Advantages of Federated Learning?}

The main advantage of FL, arising from its decentralized and collaborative learning properties, is the preservation of privacy during the model training process. Since only local model updates are sent to the aggregator agent, the data used to train the local models remains with the local node. The fact that neither the aggregator agent nor the other federated nodes have access to a given node’s individual data unlocks an incredible potential for FL to be used in privacy-sensitive applications. One such privacy-sensitive application which has already begun to implement FL is the intelligent healthcare sector \cite{FF1} . The most prevalent example in this sector is collaborative model training between hospitals. Traditionally, due to data privacy and patient confidentiality concerns, hospitals had only had access to their local patient data when training models, which created a data-availability bottleneck and was a key deterrent for the use of ML since the local data was insufficient. With FL, several hospitals can contribute to the training of a global model, which leverages updates from all their local models, thereby enabling an individual hospital to use insights from other participating hospitals to improve the performance of their models. This collaboration leads to the second main advantage of FL, data availability. Through the use of FL, not only is it possible to ensure data privacy, but it is also possible to use insights from non-local data, thereby increasing the total amount of data used for model training. This is especially important for nodes with low amounts of local data as an inadequate amount of data can prove to be detrimental to a model’s performance. Additionally, the method by which this data is processed is extremely advantageous as all nodes are solely responsible for processing their local data. The ability to leverage global data but only process local data makes FL especially applicable to resource-constrained nodes with low processing capabilities, such as those found in MEC IoV systems.\par

 Another advantage of FL is the communication efficiencies experienced during the model training. Since there is no data transfer between local nodes and the aggregation agent, and no communication between nodes, FL is very efficient compared to centralized collaborative learning strategies. When considering the complex IoV and ITS network layout, this communication efficiency can alleviate a potential strain on network resources during times of high traffic, something which is increasingly important considering the critical services using the network and its services. \par

 The final key advantage of FL is seen in a fault or failure mitigation scenario. In the case that a local node goes offline or loses connectivity, the FL training process can still continue normally. In the case where a node loses connectivity, it still has its local model to use and, therefore, won’t experience a fault. However, in the case where a node goes offline and loses its local model, once communication is restored, the aggregator agent can quickly push the current global model to the node. Having a global model the node can resume operation and restore performance. Furthermore, having a global model and then applying local training is an incredibly efficient failure mitigation strategy. Firstly, the model has already been developed and trained through a number of iterations. Secondly, local data from the failed node contributed to model development before the failure. Considering the abovementioned points, as soon as the node receives the global mode, it can begin using it, and through a minimal number of local training iterations, pre-failure performance can be quickly restored. The effectiveness of this mitigation strategy is highlighted when compared to alternative schemes, such as complete model retraining and the reinstatement of historical models. The complete model retraining strategy, as the name suggests, requires the entire retraining of the mode from scratchl using local data, something which is very time efficient and can increase downtime. The historical model reinstatement firstly requires the storage of previous models, and secondly, the retraining of these models, which is an improvement to the complete model retraining strategy, however, can prove to be inefficient in a highly dynamic and volatile network scenario. Since the global model is ‘live’ and constantly updating based on incoming data, it is the preferred failure mitigation strategy. To illustrate the capabilities of FL and its advantage over other failure strategies, we present a case study.

\subsection{Case Study}
\begin{figure*}[ht]
\centerline{\includegraphics[width=30pc]{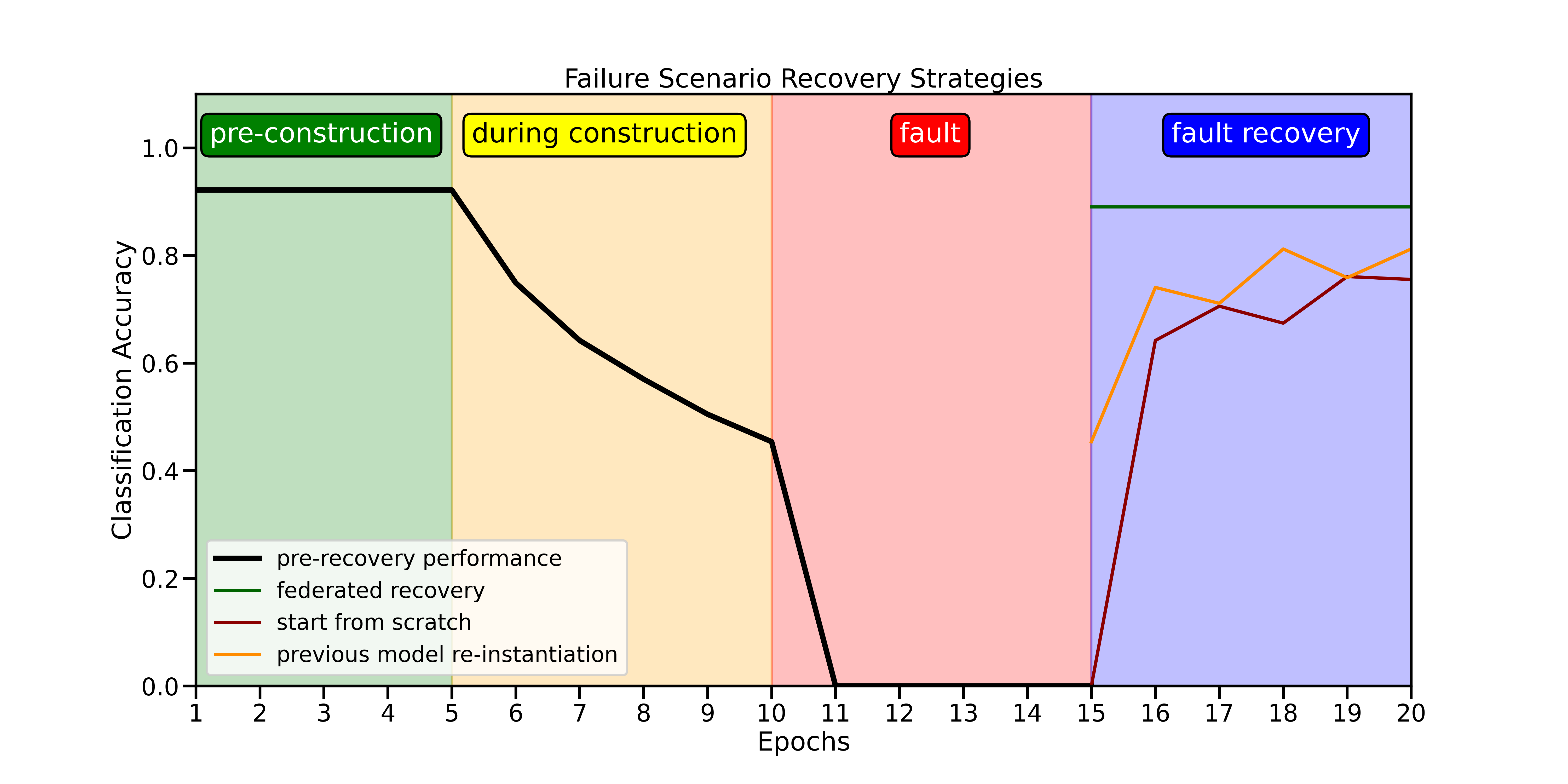}}
\caption{Failure Recovery Strategy Comparisson}
\end{figure*}
\textit{A Federated Learning – enabled RSU is located in a low traffic area and performs object classification tasks. Due to a recent construction project, the neighbourhood is experiencing a variety of previously unseen traffic. The RSU continually trains its local model to adapt to the changing environment; however, it experiences an outage and loses its local model. It can select one of the previously outline failure recovery strategies. Which one should it choose?} \par

To simulate the scenario described above, the MNIST digit dataset \cite{GG1} was used. Initially, a model is trained on a set number of digits ({\it i.e.,} even), which simulates the operation of the RSU before the construction project. Once the construction project begins, and previously unseen traffic is experienced, odd digits are gradually introduced into the training and testing datasets; however, training of this model is not complete as the fault is experienced. There are three possible fault recovery scenarios outlined: \par
1.	A model is retrained from scratch using the local data available. The training and testing sets of this model will consist of even and odd integers to simulate the moment at which the fault was experienced. \par
2.	A previous model is re-instated, and then training occurs using local data. The dataset used to train this model is the same as scenario 1. \par
3.	A global federated model trained using 10 system nodes representing RSUs spread throughout the network is pushed to the failed node, and is instantly used to resume performance.

A comparison between the three mentioned failure mitigation strategies is exhibited in Figure 4. As seen through this figure, the federated recovery strategy performs the best as it instantly restores normal operation and performance returns to acceptable levels without any additional training iterations.\footnote[1]{code for this use case is available at \url{https://github.com/Western-OC2-Lab/FL-IOV-ITS.git}}
The main advantages of FL described throughout this section encompass many operational benefits, which are summarized in Table 1.

\section{Applications}

The following section outlines some of the various applications of FL in the IoV and ITS, including RSU Intelligence, NFV Orchestration, and Vehicular Intelligence.

\subsection{RSU Intelligence}

As previously mentioned, ITS have entities known as RSUs, which will be located along roadways. These RSUs will be equipped with sensors to collect data and will possess basic processing capabilities. When considering the general FL application architectures, RSUs perfectly match since they are repeated entities capable of data collection and processing. Since the RSUs will be collecting (and receiving) a variety of different data, they are capable of applying FL in many different scenarios. One of the most important and prevalent applications is image processing. When considering a system of fully autonomous vehicles, image processing is essential both onboard the vehicular clients as well as roadside through the RSUs. Image processing tasks on these entities can range from pedestrian detection to collision reporting. When considering any transportation system, there are differing levels of vehicular traffic experienced throughout the system; this traffic imbalance can put certain RSUs at a great disadvantage as the data collection in low traffic areas will minimal compared to the data collection in high traffic areas. To mitigate this, FL applied across all RSUs can enable the collective use of all RSU data and the distribution of a complete model to RSUs which otherwise wouldn’t have an adequate amount of data to train a local model of their own. Another advantage of using FL for image processing is the ability to leverage the differing conditions ({\it i.e.,} lighting) in RSU-collected images. Some RSUs will be placed in fully lit areas, while others might be placed in neighbourhoods with many trees and shaded regions.  To ensure proper object detection, all RSUs should have models capable of operating under varying and changing conditions. Considering the criticality of object detection applications in RSUs, FL is essential for ensuring model performance and, subsequently, the safety of drivers and pedestrians alike.

\subsection{NFV Orchestration}

NFV Orchestration presents a very appealing and intriguing application for FL. The most enticing property of FL, making it a candidate for NFV Orchestration is its privacy and security due to the critical services ({\it i.e.,} financial, emergency) NFV-enabled networks support. When considering the IoV and ITS as an extension of current networks, NFV Orchestration will play a critical role in vehicular service delivery. However, there is one major difference between traditional NFV Orchestration and NFV Orchestration for MEC enabled IoV and ITS. With many delay-sensitive applications ranging from vehicular safety and routing to immersive virtual reality services, NSPs are facing an unprecedented challenge of reducing end-to-end application latencies to less than 1ms for ultra-low latency applications \cite{HH1}. To do this, points of presence such as RSUs which are placed at the very edge of the network will act as network nodes capable of hosting VNFs.  However, without proper Orchestration, the stringent latency requirements will not be met. Traditionally, NSPs have had to resort to using near-optimal heuristic solutions to address the infeasibility of optimization problem formulations due to their runtime complexity, however, with the adoption of ultra-low latency applications, near-optimal heuristic solutions are also becoming infeasible. To mitigate the infeasibility of traditional NFV Orchestration techniques, NSPs have been exploring ML as an alternative.  Already, ML techniques are being applied to NFV Orchestration tasks such as VNF placement \cite{MR1, DD1} and migration \cite{II1} with incredible success due to their ability to approach the performance of the optimization model formulations with an incredible reduction in time-complexity. \par

When considering ML in the IoV and ITS, the use of FL is the next logical step. Due to the increasing types of services, each with its own complexities coupled with the expansion of traditional networks to MEC enabled networks, NFV Orchestration activities will drastically change. The transition to MEC-enabled networks will usher in the vast expansion of network nodes. This will require the partitioning of current networks into much smaller sub-networks, each with their own NFV Orchestrator. Additionally, incorporating aspects from 5G networking, each of these sub-networks may be further partitioned using network slicing techniques. This increased complexity promotes the use of FL as NFV Orchestrators from different network partitions can use collaborative ML training to create models capable of completing orchestration tasks such as VNF Placement, Scaling, Termination and Migration. Figure 5 illustrates the complexity of vehicular requests in a 5G-enabled ITS.

\begin{figure*}
\centerline{\includegraphics[width=25pc]{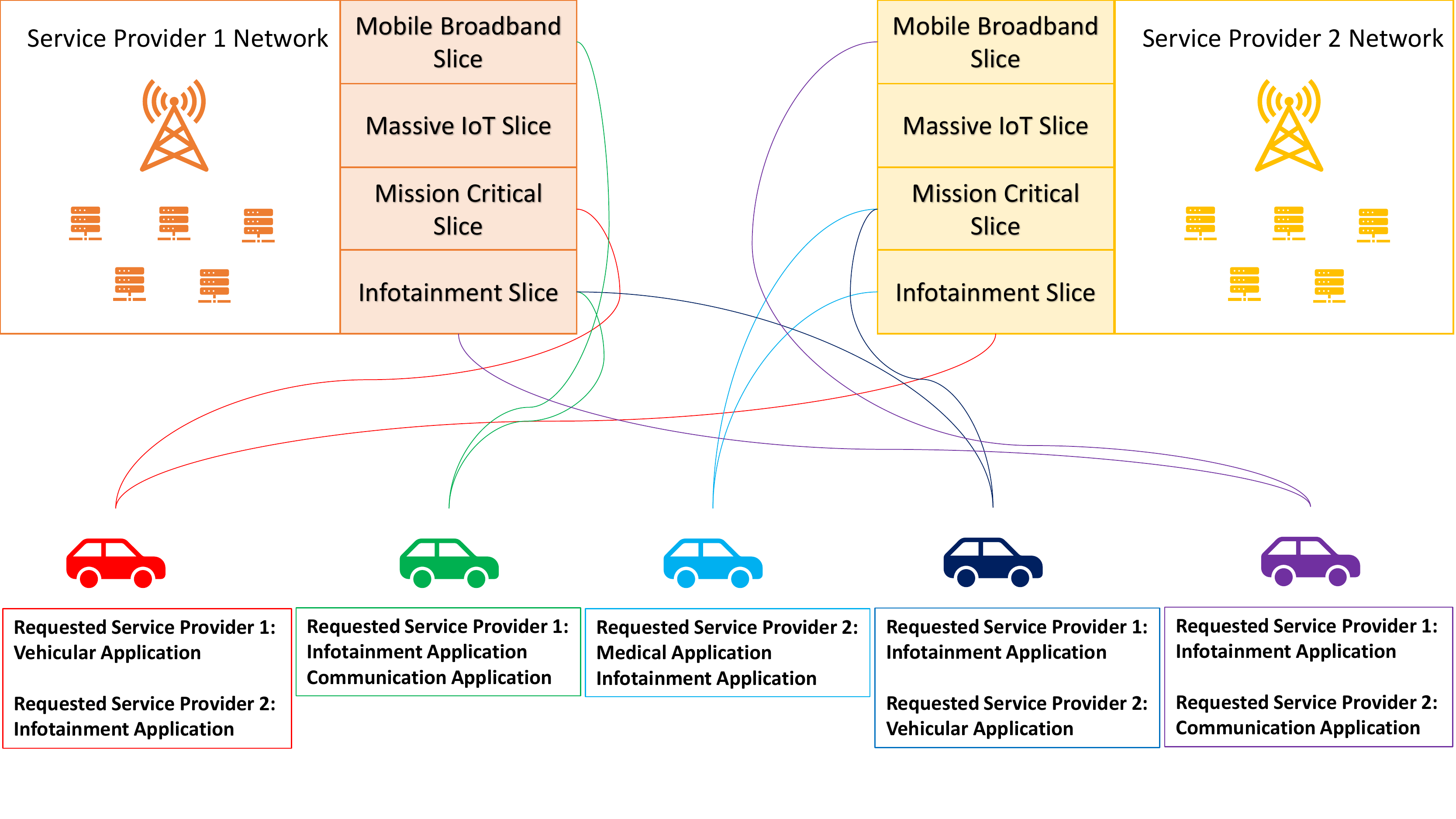}}
\caption{Vehicular Application Requests in 5G-enabled ITS}
\end{figure*}

\subsection{Vehicular Intelligence}
Vehicular intelligence in the IoV and ITS can describe a plethora of applications, including in-vehicle intelligence ({\it i.e.,} communications), image processing ({\it i.e.,} lane detection), and forecasting ({\it i.e.,} road conditions); however, a very interesting use case currently being heavily explored in the manufacturing sector relates to predictive maintenance. Predictive maintenance uses operational data to predict when a specific component will fail and suggests a maintenance schedule that will pro-actively address the failure through planned maintenance. Unplanned maintenance is extremely costly in manufacturing as a break in production can result in the loss of millions in revenue. Additionally, reactive maintenance takes a large toll on assets and reduces their lifespan. Several implementations of predictive maintenance models have shown incredible success; one main example stemming from the oil and gas industry is Royal Shell Corp who has prevented millions of dollars in lost revenue and damages while improving the longevity of their assets \cite{JJ1}. The same benefits can be experienced when considering the application of predictive maintenance on vehicular clients in the IoV and ITS. It is estimated that in 2016 there were 1.32B vehicles globally \cite{KK1}, a number which is expected to grow exponentially in the next decades. In an ITS scenario, each of these vehicles would have an extensive log capturing real-time measurements from its various sensors. By leveraging FL, along with this vast collection of data, comprehensive predictive maintenance models can be built. While the mechanics of how to select which data is used to build these models is still an open question ({\it i.e.,} brand-based, location-based, network-based), the impact FL will have on predictive vehicular maintenance is indisputable. 

\section{Summary and Conclusion}

As demonstrated throughout this paper, Federated Learning has incredible potential in terms of its applicability to the Internet of Vehicles and Intelligent Transportation Systems. From the various number of use cases, including Roadside Unit Intelligence, Network Function Virtualization Management and Orchestration, and Vehicular Intelligence to the incredible number of benefits it provides, Federated Learning is a key enabler for next-generation networking technologies.

\end{document}